*Attilio Sacripanti•*^*
°ENEA (National Agency for Environment Technological Innovation and Energy)
*University of Rome II "Tor Vergata" Italy
^ FIJLKAM Italian Judo Wrestling and Karate Federation

*Judo Match Analysis a powerful coaching tool : basic and advanced tools in a fighting style evolution*

Abstract
In this second paper on match analysis, we analyze in deep the competition steps showing the evolution of this tool at National Federation level.
On the basis of our first classification, Match Analysis is a valuable source of four levels of information:
1st.        Athlete's Physiological data
2nd.       Athlete's Technical data
3rd.        Athlete's Strategically data
 4th.       Adversary's Scouting
Furthermore, it is the most important source of technical assessment.
Studying competition with this tool is essential for the coaches because they can obtain useful information for their coaching.
Match Analysis is today the master key in situation sports like Judo, to help in useful way the difficult task of coach or best for National or Olympic coaching equips.
In this paper it is presented a deeper study of the judo competitions at high level both from the male and female point of view, explaining at light of biomechanics, not only the throws evolution in time ( introduction of Innovative and Chaotic techniques) but also the evolution of fighting style in these high level competitions, both connected with the grow of this Olympic Sport in the Word Arena ( today 199 countries are members of IJF)
It is shown how new interesting ways are opened by this powerful coaching tool, very useful for National team technical management.
In the last part of this paper we analyze advanced mathematical tools describing Couple of Athletes motion as Fractal Poisson Point Processes based on Fractional Brownian Motion
to show how strategic evaluation , probability and short term forecast can be applied to
Judo competition.

# Attilio Sacripanti °*^

°ENEA (National Agency for Environment Technological Innovation and Energy)
*University of Rome II "Tor Vergata" Italy
^ FIJLKAM Italian Judo Wrestling and Karate
Federation

Judo Pictures IJF Archive Courtesy by IJF President

# Ju Do Match Analysis a powerful coaching tool
*(Basic and advanced tools in a fighting style evolution)*

## 1) Introduction

In the last 54 years judo fighting style and biomechanics studies, starting already from Japan, walked together along the world's tatamis [1,2,3,4...10]. In Judo world, many people speaks about changes and evolution/involution in judo competitions, this last argument comes both from the change in the evaluations criteria, the refereeing rules and the introduction over the year of different (new?) judo techniques.

If we make a comparison among Judo and other Olympic Sports, it is easy to see that Judo is already on the way to find the best rules for competition apt to underline its beautiful and esthetic dynamics, connected to throwing techniques.

Judo in the mind of his founder Jigoro Kano, was an educational system able to teach also the esthetic sense of disciples ( for example a throw who obtains Ippon; or a beautiful Kata performance ), in this aspect throwing technique is the research of maximum effectiveness respect to the effort applied. However the esthetic aspect of judo (during performance) is today not well enhanced and underlined by referee rules.

In this paper, we will speak about the evolution of judo fighting style from the technical point of view, by match analysis evaluation, introducing some special considerations that are able to improve the general forecasting and analysis capability of this advanced tool.

Today the Continental and National Federations under the International Judo Federation (IJF) directives, are pushed to use this special tool in centralized way, to save money rationalizing software and methods .

EJU utilizes a specific software , equal for all countries. This software for example, in off line analysis, just at the end of each competition performs a very first centralized gross elaboration of fighting data and these data are sent to all countries, after that each country develops a personalized analysis in his own way.

## 2) How to analyze judo competitions.

### 2.1 Technical Steps in competition

It used by most National Federations' analysts to divide the whole competition in technical steps:
The basic step in the time domain is not fixed, because the analysts normally use different technical finding to analyze the competition, normally there are used five or six technical steps.

*A)* In our paper we present first the subdivision in six technical steps ( Francini and coworkers) [11]:
1) *Matte* – a period of interruption, when the referee calls *Matte* to discontinue any activity of the judoka;
2) *Preparation* – a period of movement, observation and non-contact preparation;

3) **Grip** – a period of the match in which the judoka disputes for the best grip (*Kumi-kata*), when there is contact with one or both hands;
4) **Throws** – a period in which the judoka executes a technique or throw during standing combat (*Tachi-waza*);
5) **Fall** – the moment of the fall (*Kake*) when the analyzed athlete falls to the ground being possible that both judokas fall as a result of the application of *technique*.
6) **Groundwork** – any combat that takes place on the ground (*Ne-waza*), when strangle or arm-lock techniques are applied with the aim of immobilizing or finalizing an opponent.

***B)*** The second model comes from England is the Hajime –Matte model proposed among other by Karen Roberts [12] always in six phases with the following key points:
1) ***Mobility (shifting alone)*** - Dynamic Posture and Stance, Distance, Tsugi-Ashi, Tai-Sabaki, Ashi-Sabakai, Tandoku-Renshu (Coordination)
2) ***Kumi Kata*** - Lead Grip / Main Grip , Structure, 'First On' Getting Your Grip, Hiki-Te, Tsuri-Te, Tsugi-Ashi, Ashi-Sabaki
3) ***Preparation*** - Action / Reaction, Direction, Timing / Distance, Ashi-Waza Combinations, Feint Attacks, Kuzushi, Hiki-Dashi.
4) ***Nage waza*** - Tai-Sabaki, Speed of Entry, Transfer of Power by collision, Control, Belief, Execution, Commitment to Finish Attack
5) ***Transition*** - Mobility / Agility, 'Catch', Positioning, Continuous Control, Dominate, 'Open Up' Opponent
6) ***Ne waza*** - Mobility, Control, Connection, Multipurpose Attack, Belief

It easy to understand that competition is a linear connection of these steps.

$$C = \sum_{i=1}^{6} \sum_{k=1}^{N} \frac{\mu_{ki}}{1}\left(\tau_1^{\beta_1} + \tau_2^{\beta_2} + \tau_3^{\beta_3} + \tau_4^{\beta_4} + \tau_5^{\beta_5} + \tau_6^{\beta_6}\right)$$

In which is one phase μ the number of the specific phases in each completion, and is the existence factor of such phase in each competition = 0 or 1 (yes or not).
In Judo, the distance into the Athletes couple depends both from the Kumi-kata and the feet position, things that determine the guard position. These two parameters change many times during competition and in very important way during attack.
"Rigid" grips are negative for the distance variations, and they are wrong for the multiple attack capabilities.
The special Kumi kata can be considered psychologically encouraging, but from one other side it not lets to attack in many different directions.
The basic Ki hon kumi kata lets to apply more attacks in different directions and it can be considered more generalist and preferable.
Analytical studies on the *Kumi Kata*, used in competition, like right, left, mixed *Kumi Kata*, studied in Japan (*Analytical studies on the contests performed at the all Japan Championship Tournament*) [13]. If we think deeper, the motion on the mat, it is the results of many pushes and pulls applied by the grips, but it is not possible to apply push or pull without the contact to the mat by the feet. Thinks for example to apply the same push-pull forces in the couple system, wearing roller-skates, then obviously it will be impossible to apply anything!
Now after that, it is understandable the meaning of the so called

"***Biomechanics Grips Paradox***".
*What is the most important aspect of the Grips (Kumi Kata)?*
*Feet Position is the Grips most important aspect!*

Arms position is essential only in defining the forces' directions to throw the adversary, but without a strong support base the arms position is unimportant or unuseful.

This is a new kind of fight vision, not to see simply in short way: or the arms position, or the bodies' relative position, or the power applied to the adversary.

But the advanced way is to approach the system as whole seeing at the Couple of Athletes and not at the single athlete.

This new vision is the right biomechanical vision, or in other words the advanced modern vision of the Judo competition.

This aspect, very important, from the strategic point of view, as seen, came from the study not of the single athlete, that coach normally perform during competition, but from the analysis of the whole system "couple of athletes".

If the competition is approached in this way many interesting adversary's aspects, both from the strategic and technical (Throwing) point of view get out from the system observation.

and the final result is very useful for a right high performance coaching.

## *2.2 Competition Invariants*

The ***Competitive Invariants*** singled out by the author are the so called "Guard Position" which are the hold or grip positions that the couple of athletes closed system got during the fights. ( see Biomechanical analysis of competition) [14].

These positions were classified on the basis of two relative ranges: distance between the heads, and distance between adversaries' feet in two main groups.

These groups are connected to the couple of athletes' shifting velocities, each group could be divided in three subclasses related form left to right to increasing shifting velocity of Couple. (see Fig.1)

These positions are strongly connected to the preferred fighting motion pace of each athlete, which reveals a lot of technical information about the fighting preference and the special class of Throwing utilized (Tokui Waza).

Normally during competition the normal athletes' attitude changes:
1) They lower their center of mass position, to stabilize
2) The body superior part (head and shoulders) is closer the inferior one's (hip and feet) is far
3 ) They increase the mutual control.

Generally speaking the control increase by a strong fixation of wrists, simultaneously drawing down by arms and using one's weight to slow down the adversaries' movements, but without tense the arms. The hikite arm is utilized during attack to open the adversary guard, in defense by the control of cuff jacket; it is able to stop the adversary's attack.

It is interesting to remember the golden rules of Kumi Kata stated by Kazuzo Kudo in his superb textbook Dynamic Judo:

" *Do not tense your arms, but stand as loose and flexible as possible. Not only is this nearly impossible for beginners, it is trouble-some for those men who made considerable progress in judo. The truth of the matter is that as you recklessly thrust your arms out and pull them in during practice you will naturally come to understand what the technique is all about and will master the use of your arms. When this happens you will be able to tense your arms, when you need to and relax them, when you do not. An important thing for you to understand at this stage is that you apply judo techniques as you move back and forth and from side to side together with your opponent and that you must relax your arms as you do so.*" [15]

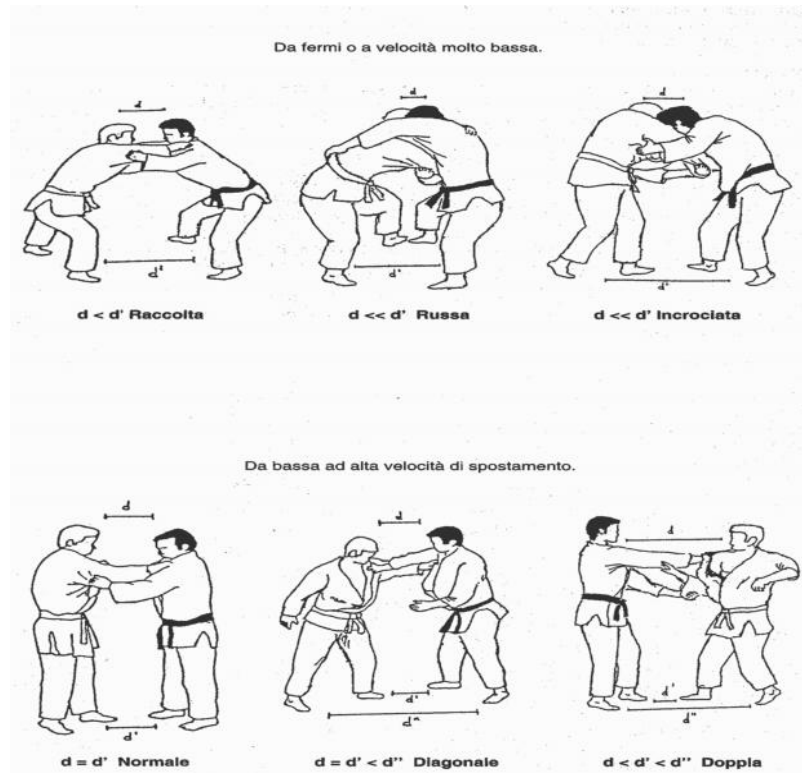

**Fig.1** *Six Classes of Guard Position (Competition Invariants) related to the couple shifting velocity. (Sacripanti)*

The six classes of "*Competitive Invariants*", fig (2-7) connected to the couple's increasing speed, in which it is possible to collect together all the infinite grip positions depending also by refereeing rules, that couple of athletes closed system could built.

These are the position classes that the couple of athletes closed system could built. Biomechanically speaking, today it is not allowed to take the belt or under the belt, each position is connected to the Tokui Waza and the pace that the Athlete likes to apply in such motion situation.

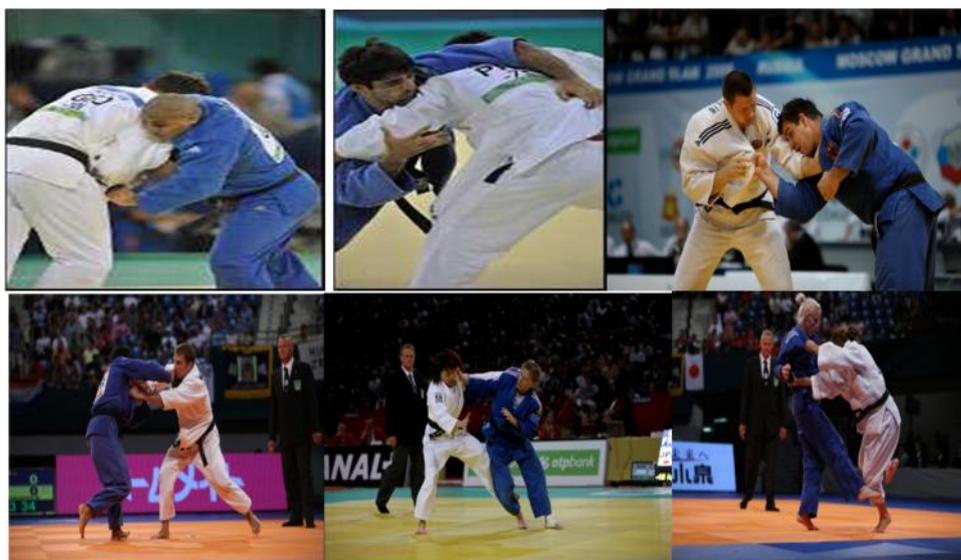

**Fig.2-7 The Six Classes of Guard Position related to the couple shifting velocity in real competition. (Finch, Zahonyi )**

Then, generally speaking, if the coach sees at the Guard Position (**Competitive Invariant**) and understand the pace motion of the adversary he can preview the biomechanical class of his preferred Tokui Waza.

Remembering that the biomechanical tools are connected to the shifting speed, or in other words that it is easier to apply Techniques of couple of forces at high shifting speed, than techniques of the physical lever, it will be easier, for example, to recommend to his own athlete to change the motion pace, in such a way as to increase the difficulty for the adversary to apply the Kuzushi-Tzukuri phase connected both with his Tokui Waza and with the preferred speed.

Today with the fight evolution elite athletes are able to change guard position during fight, by the way the connection speed throws is always valid and useful to contrast the pace motion changed.

Obviously in such situation to attempt the victory means to connect all these information in a whole strategic approach to fight.

How this information, without singling out the concept of Competition Invariants, is present in Japanese Judo it is easy to see in the following recommendations of Sato and Okano in "Vital Judo" about Uchi Mata, they present five methods to solve the Kuzushi-Tzukuri phases starting when a small Tori attacks Uke that is in traditional grip in Ai Yotsu, two methods when the opponent stands with both arms rigidly out-stretched used mainly by Tomio Sasahara, one when the opponent has retracted his hips and assumed a position of stubborn resistance, the last favorite method of Hirobumi Matsuda.

In these differentiated attacks connected to the Uke's body position there is the ground of both concepts: Couple of Athletes System and Competition Invariants. [16]

### 3) The Four Levels of Match Analysis

In these times, Match Analysis was developed in preferential way, for team sports mainly football, soccer, basketball, volleyball, and these hardware and software complex are today at the up data of the sport's world.

For dual Situation Sports like Fencing, Tennis, Judo, Karate and so on [17-20]

There are today two main utilization of video data collection:

**a.** *In Real Time*
**b.** *Off line*

The most used software for real time application come from volleyball, but also all other sports team have real time applications, dual sports use match analysis systems mainly in off line studies. However we focalize ourselves on Judo application.

*Real Time*

For real time application judo as dual situation sport with short time technical executions is a difficult task for software program, the only application is to help coach to give to the athlete the right and effective strategic or tactical suggestion, depending from the fighting style of the adversary. However remembering that the technical solution of whatever fighting situation in judo, could be also an irrational or chaotic solution, outside the logical development but precisely for this reason very effective.

This give us the limit of real time application of Match Analysis in real competitions.

More often the real time application in judo is more useful for refereeing evaluation and/or correction.

*Off line*

The off line application of judo is more widely used. One other objective reason is that judo match is short in time and this situation makes more difficult the utilization of real time analysis.
Technically speaking data video obtained will be quantitatively useful only if the camera is calibrated. In other way, only qualitative analysis can be performed.

The most important problem, in the off line analysis case, is focalized in the saving time automatic treatment of data base content.
On the basis of the technological development, quantitative or pseudo-quantitative Match Analysis could be today a valuable source of four levels of coaching information.
These new interesting ways opened by this powerful coaching assist, are very useful for all judo operators.

*1st. Athlete's Physiological data*
*2nd. Athlete's Technical data*
*3rd. Athlete's Strategic data*
*4th. Adversary's Scouting*

### 3.1) First Level Physiologic Data

In the Area of Physiological data, the quantitative Match analysis, well performed, could help in three main ways:

a. Gross evaluation of the more precise energy cost as input for conditioning
b. Control of conditioning development and effectiveness
c. Accident prevention

Normally energy cost could be connected to the time structure of the contest.
Several researchers made evaluation on this field and the most important results are shown in the following table.

| Authors | Activity (s) | Pause (s) |
|---|---|---|
| Castarlenas and Planas (1997) | 18 ∓ 9 | 12 ∓ 4 |
| Monteiro (1995) | | |
| 1st min | 25.8 ∓ 7.8 | 9.5 ∓ 3.2 |
| 2nd min | 27.0 ∓ 9.0 | 10.4 ∓ 4.5 |
| 3rd min | 27.0 ∓ 9.7 | 13.4 ∓ 7.6 |
| 4th min | 22.4 ∓ 9.3 | 13.2 ∓ 7.3 |
| 5th min | 18.9 ∓ 10.4 | 13.9 ∓ 9.0 |
| Sikorski et al. (1987) | 30 | 13 |
| Sterkowicz and Maslej (1998) | 25 | 10 |
| Van Malderen et al. (2006) | | |
| Female | 19.9 ± 7.3 | 7.5 ± 6.2 |
| Male | 18.8 ± 9.0 | 9.1 ± 5.1 |

*Tab.1 Time Activity in fights*

Although studies on the physiological demands of judo have previously been performed, some new perspectives were founded by Degoutte et alt. [21] because the energy requirements were evaluated during a judo match, and not in a laboratory. or in special exercises.
The evaluation was performed by nutritional approach and they show that a judo match induces both protein and lipid metabolism even if the anaerobic system is brought into action, with mean levels of plasma lactate of 12.3 mmol/l.
The meaning of this results was that glycogen in the muscle was not the only substrate used during a judo match.

Several factors such as carbohydrate availability, training adaptation, and metabolic stress must be account for the use of these substrates.
In the following figure (fig 8) it is possible to see one example of first level results of Match Analysis applied to the phenomenal champion Teddy Riner from France  [ 22]

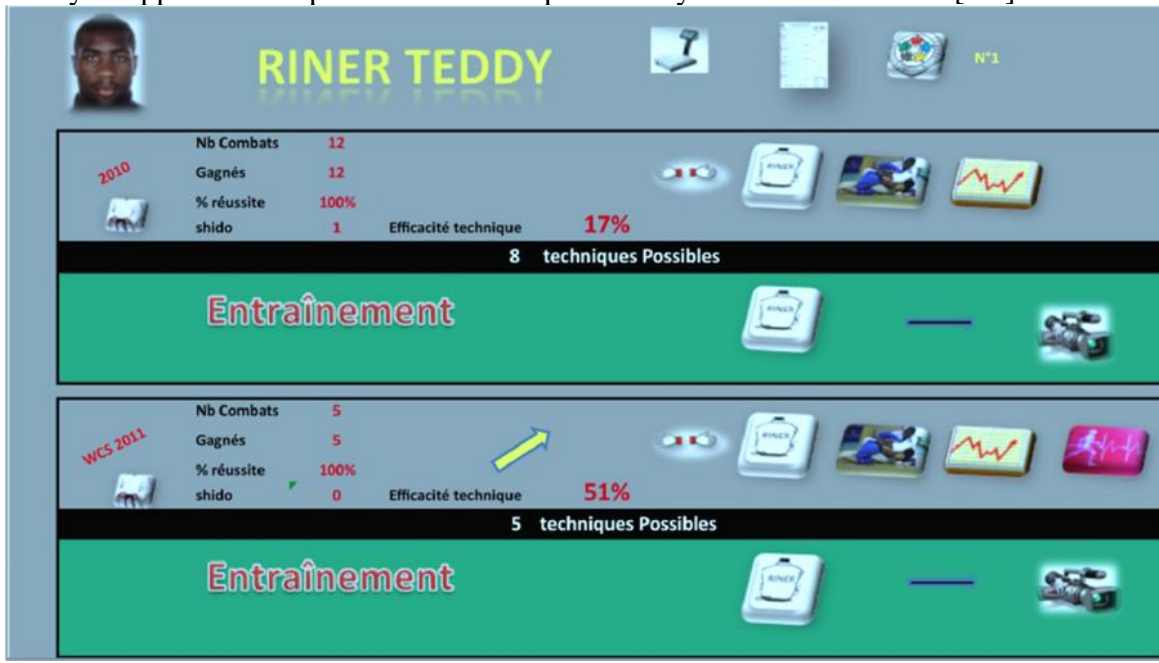

**Fig 8  Results of first level Match Analysis in the Judo French Federation [22]**

*3.2) Second Level Biomechanical technical improvement*

In the area of the Technical data, essentially qualitative Match Analysis could be a very useful source of information for the athletes' biomechanical enhancement, in six main ways:
a. Biomechanical evaluation of grips methods and fight.
b. Biomechanical improvement of throws techniques.
c. Biomechanical improvement of Standing-laying connection
d. Biomechanical improvement of Ne waza techniques.
e. Biomechanical improvement of defensive systems.
f. Weak point's analysis.

In the following figure (fig 9) it is possible to see one example of second level results of Match Analysis applied to the phenomenal champion Teddy Riner from France  [ 22]

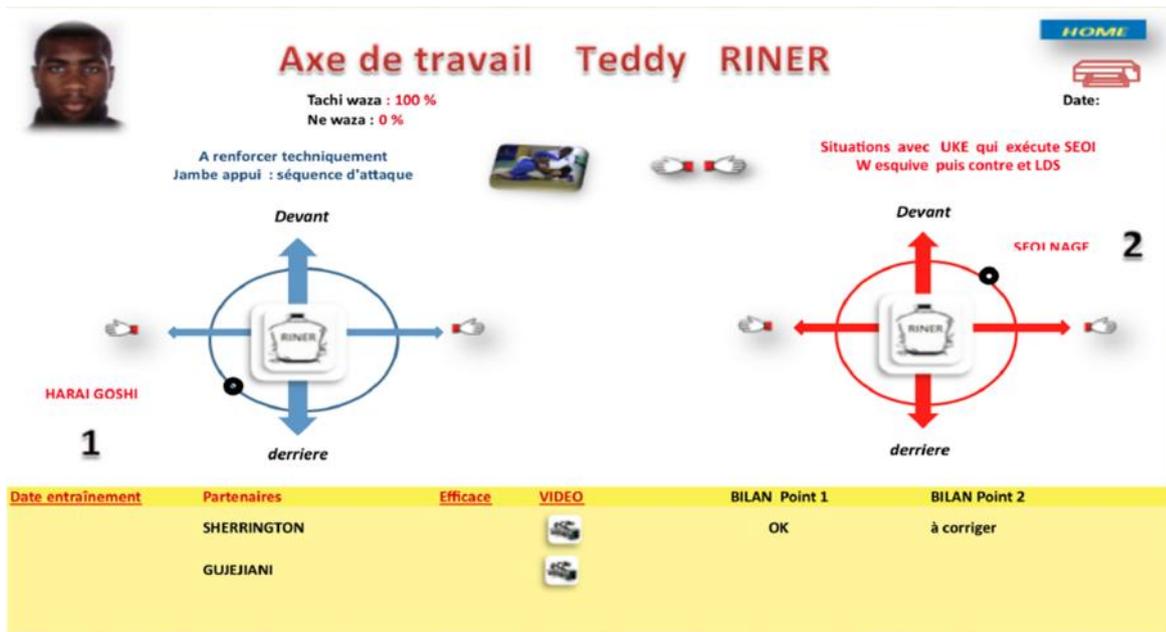

**Fig 9  Results of second level Match Analysis in the Judo French Federation [22]**

### 3.3) Third level Strategic Data

In the area of Strategic data, qualitative-quantitative Match Analysis could be a very important coaching assist, in the study and the refinement of strategic plans adopted; by two ways:
a. Local Strategies.
   Local strategies are connected to the technical-tactical solution of special situations that can occur in a specific small areas of competition,  for example action-reaction trick in the corner area, and so on.
b. Global Strategies.
   Global strategies are focalized on the whole competition contest, : way to fight when in vantage , way to fight in disadvantage,  special strategy for the golden score, penalties management, etc.

### 3.4) Fourth Level Adversary scouting

The same qualitative data base of athletes could be studied from a different point of view, to gather information about most dangerous adversaries for each fighter, by the main utilization today:
 Adversary's scouting.

In the following figure (fig 10) it is possible to see one example of fourth level results of Match Analysis ( Adversary Scouting) applied to the Italian  champion Elio Verde from the Judo French Federation  [ 22]

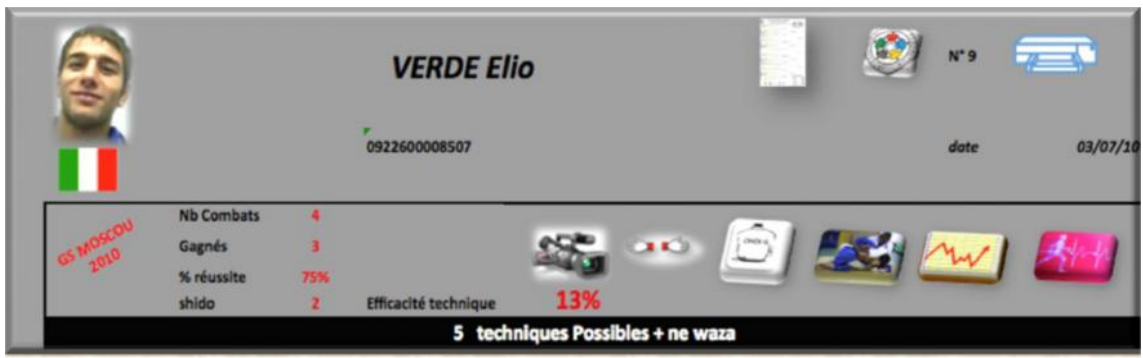

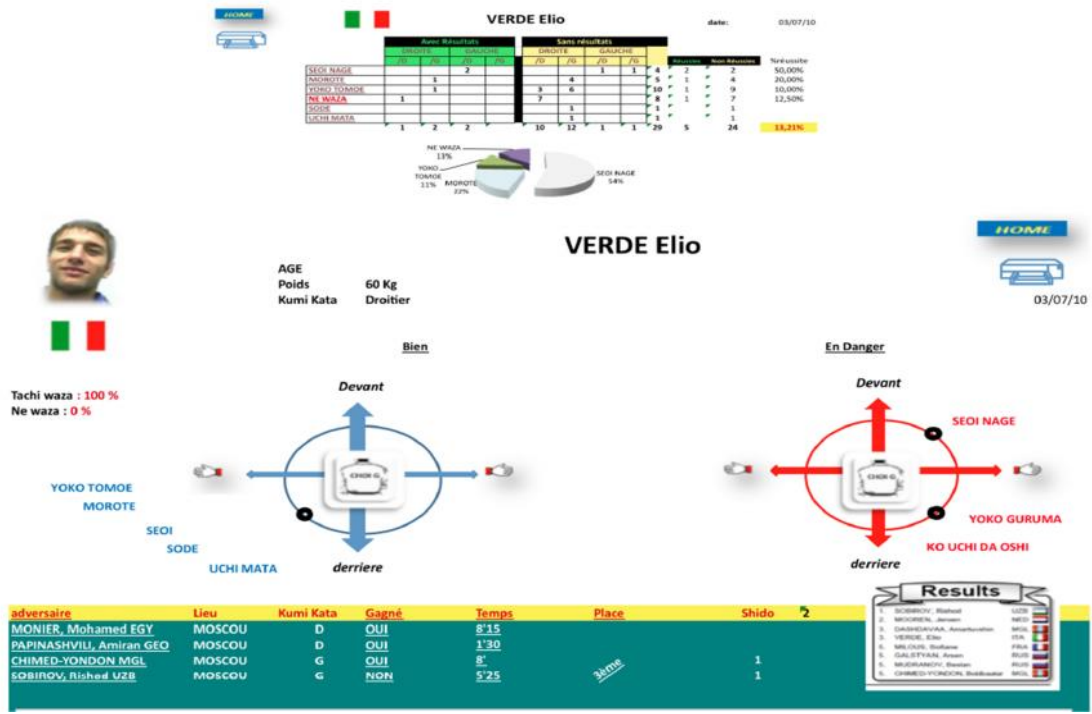

**Fig 10 Results of Fourth level Match Analysis ( Scouting) for the Italian Champion Elio Verde, made by the French Federation, see how specific and useful are the information obtained [22]**

The previous results fig.( 8-10) are a clear example of National Federation utilization and analysis of centralized data obtained from the EJU.
It is interesting to note that for the athletes scouted it is possible to know the strong and weak points in term of throwing techniques applied and suffered during the competition.
More often such strong and weak points are connected also to spatial directions respect to the body, to the grips preferred, to the effectiveness of the attack and so on.
EJU utilizes as before remembered a specific software, equal for all countries. This software, in off line analysis, just at the end of each competition is elaborated in one gross elaboration of fighting data and these results are sent to all countries, after that each country develops a personalized analysis in his own way, like the results presented in the previous figures.
Normally National Federation use software to analyze competition performance, create performance data, skill development, profile opposition and provide live video feedback within technical training sessions.
the software utilized is easy to use and a very effective coaching tool, the last development provide also a TV connected software that has proven to be an invaluable and cost effective online video system, which means that Federation can provide video data quickly to coaches and athletes where ever they may be in the world. With the development of IPhone; IPad apps, it is providing the needs of the Performance Analyst to create complete package useful to meet the modern requests of high performance sport like judo.

### 4) What are the most important biomechanical parameters in competition?

If we think at competition analysis it is well known that each competition is single and it is impossible to repeat one competition exactly, because in term of advanced biomechanics, judo competition could be defined as:

***A complex nonlinear system in which it is not possible to find a repeatable motion pattern in each game, and for each game, both motion of the Athletes Couple and Time Attack Pattern are random processes.***

If Judo competition is a stochastic process we need to analyze it of some invariant or recurrent situation equal in every contest.
What is equal in each contest? Obviously Athletes Couple that are characterized by their grip position! But grips are not only the arms position but the whole body position considering the complete connection among bio-kinetics chains.
In that way born the necessity to study the previous introduced: ***competition invarian**t*.
 The correct biomechanical way to analyze such macro phenomena is to study them. in two steps:
One to study the motion of the couple, two to study the interaction into the couple ( throwing techniques).
 The analysis of "Couple of Athletes System", during Judo fight, single out that, at light of Biomechanics, the basic parameters to perform effectively are, only three.

**$a_1$) – *Shifting Velocity***

**$a_2$) – *Attack Speed against Reaction Capability***

**$a_3$) – *Bodies' Relative Positioning Management***

**$a_1$) – *Shifting Velocity*.**

The "Couple of Athletes" shifting velocity on the Tatami is the speed of the couple system, seen as a whole. There is important, because at every velocity class (low, average, and high) it will be possible to apply in easy way some specific preferred standing techniques.
From that it is clear that it should be possible to manage a specific competition strategy, because every competitor has his own preferred pace of shifting velocity, depending from his own Tokui Waza.
 Therefore if one athlete shall be able to compel the adversary at another disliked shifting velocity, it will cause to him a technical-psychological trouble.

**$a_2$) – *Attack Speed against Reaction Capability*.**

The very effective attack speed, to be tactically effective, must be, as high as possible.
This speed is a physical-technical athletes' capability and it is a capability able to be trained.
It is essential to remember that increasing in attack speed; it must not be to the detriment of precision of technical gesture which should be very flexible to fit the infinite and possible fight situations.

Then there is advisable before to increase the speed and after to better the precision, with the growing of his own technical maturity.
Judo, attack speed is certainly, in most of cases, shifting velocity independent, but it is also the most important parameter connected to the throwing success.

But at high shifting velocity of "Couple", fast attacking is connected on high "coordinative" skill. The increase of attack speed over his own personal maximum is very complex or impossible, but there is a very interesting biomechanical tool, to prevent the competitor reaction, the decrease of "*attack steps*" and attack "*logical path* "
Both "*attack steps*" and attack "*logical path* " are connected to shortening: distance between athletes, the right position in which one athlete can throw the opponent with the minimum energy waste.
Obviously, these movements are infinite in number; however, they pursue one common and definite objective: **to shorten the mutual distance.**

Indeed, this is a common aspect of the infinite number of situations that might arise [11,12]. Biomechanical analysis of this aspect shows some very interesting properties. It turns out that there are in fact only three classes of actions (trajectories of movements) that at the same time involve minimal energy and strive to achieve minimal distance.
In jūdō, that what we term *Action Invariant* refers to the minimal path, in time (like the Fermat principles in optics) of the body's shift, necessary to acquire the best *kuzushi* and *tsukuri* position for every jūdō throw.
Conversely, in those cases where it is actually possible to identify such a minimum action, or *Action Invariant* , the two following biomechanical axioms, that are the scientific translation of Kano's minimum energy expenditure principle, could be defined:

a) *Most effective is the Judo Technique, minimum is the Athletes' energy consumption.*
b) *Most effective is the Judo Technique, minimum is the Athletes' trajectory for positioning.*

As previously explained, we call these movement classes *General Action Invariants (GAI).* This term covers the whole range of body movements intended to both reduce the distance between both opponents and to optimize one's body relative to the adversary's body position.
Similar class of movements were found by S. Sterkowicz and coworkers in experimental study on competition: analyzing differences in the fighting methods adopted by competitors in All Polish Judo championship [23, 24].
The analysis specifies that only three *General Action Invariants* are included in the *kuzushi-tsukuri* phase for each *Competition Invariant*, namely:
1. Reducing the distance without rotation.(straight line)
2. Reducing the distance involving complete (0°to180°) rotation clockwise/counter-clockwise.*
3. Reducing the distance involving a half (0° to 90°) rotation clockwise/counter-clockwise.*

*These angles are valid for athletes in stationary position (study), but during a competitive match they may change if the athletes are moving; a classical example of this is the Japanese concept of Hand -no-kuzushi [Lit.: "reactional/recoiling unbalancing"]*

Each of these *General Action Invariants* is simply linked to the two biomechanics tools involved in effecting a throwing technique: "Couple" or "Lever".
To increase the effectiveness of throw, in real competition, athletes' bodies must collide each other.

In effect because it is difficult if not impossible to fit in, during competition in the position that people train itself normally in *uchi komi* or *nage komi,* applying forces in classical directions, in that way borne the **Innovative Techniques** ( Roy Inmann) [25] that are defined [26]
*"Innovative Throws" are all throwing techniques that keep alive the formal aspect of classic Jūdō throws, and differ in terms of grips and direction of applied forces only. Fig11*

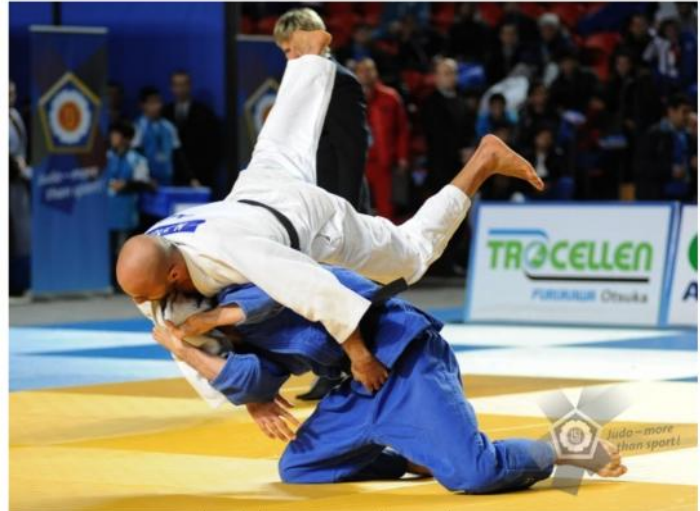
*Fig 11 Innovative Throws*

**Innovative Throws** are **henka** (variations) applications of classical *Kōdōkan* throwing techniques, which biomechanically are either Couple of Force-type or Lever-type techniques, while it remains easy to still recognize a basic traditional technique (40 *gokyō* throwing techniques) in them.

However, there are other "non classic" solutions applied in competition and which are different from 'Innovative' (*henka* Throws), which we define as "**Chaotic Forms**". Oftentimes these **Chaotic Throws** are mainly limited by the class of lever group.
When one analyses these types of throws more in depth, then the real difference between the goals of kuzushi/tsukuri in both biomechanical groups of throws will become clear.
*" Chaotic Throws" are characterized by the application of different grips positions which applying force in different (nontraditional) directions, while simultaneously applying (stopping points) in non-classical position, utilizing "no rational" shortening trajectories (longer than the usual) between athletes Fig 12*

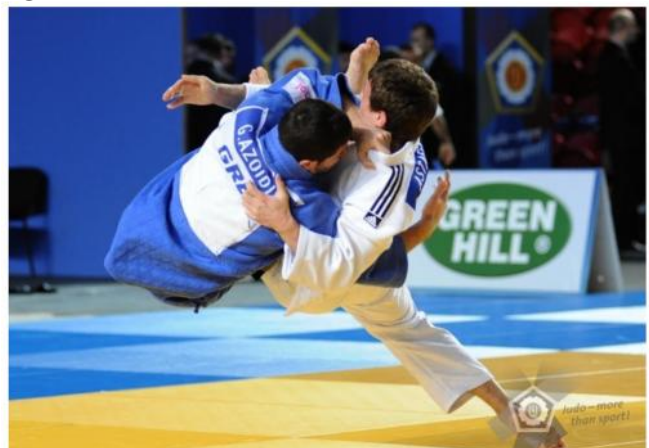
*Fig 12 Chaotic Throw*

Throwing techniques produced in such way are normally called 'chaotic', and they present high psychological difficulties in facing them.

A last consideration shortening the time to perform the General Action Invariants or applying a totally different *GAI* increases not only the throwing effectiveness but prevents also the activation of competitor *Reaction Capability*.

**a$_3$) – *Bodies' Relative Position Management.***

The capability to manage rightly the bodies' relative position is essential to the execution of a specific motor skill, like a throwing in non-conventional relative position produced by competitor reaction (Hando No Kuzushi).

5) *Throws Evolution in Time.*

On the basis of our definitions, it is clear that classic throws defined as the.

*"Classic Throws" are all throwing techniques that are explained in the world Gyms following the 40 throws ( that could be named Sportive Rational Techniques) proposed by Kano and his students. Fig13-14*

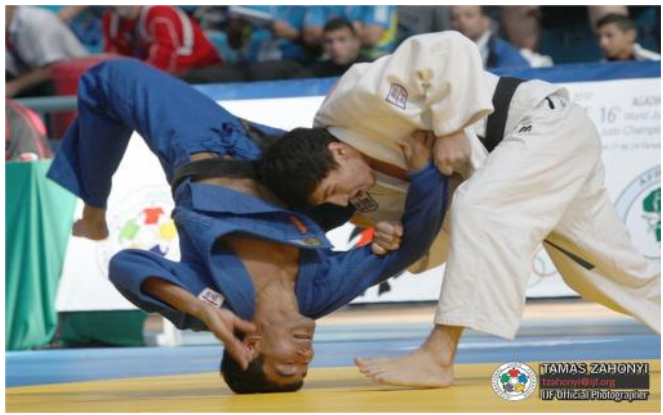

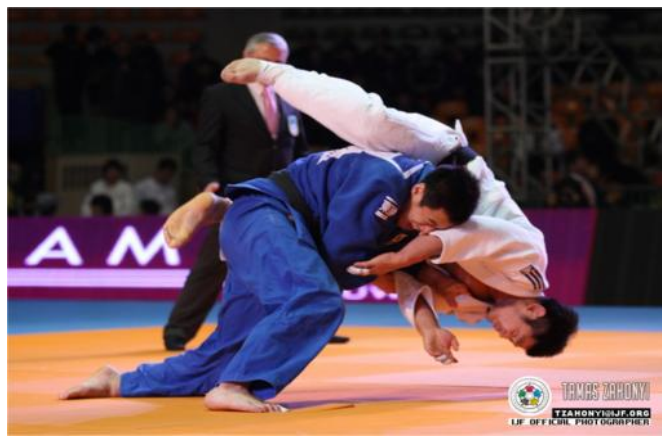

*Fig 13-14 two Classic forms of the same Throw ( Seoi Nage)*

They are a very rare event because it is very difficult to apply in real competition forces in the theoretical directions, in connection with basic grips ( Ki Hon Kumi Kata) . Normally speaking during time Judo people try to apply throws in the classic form but often the dynamics of situations needs real time fitting with the adversaries' changing positions, producing the so called Innovative throws, that are all the variations ( Henka) depending both from the high dynamics of competitions and the anthropometric differences among athletes.

From the theoretical point of view it is easy to think that more old are the competitions more easily classic throw could be applied. But also in old time henka are often present on the Japan Tatami. During time with the changing of competitions rules and the increase of number of countries, many classic grips or throwing forms were broken with the born of the well-known *Innovative variations.*
In the last forty years with the Olympic Games and the increase of countries till to the actual 199, many different ( non-classical) applications of the two biomechanical tools born in the world, with the rise of the *Chaotic throws* phenomenon , that is today more frequent than some years ago.
A further evolution from the Biomechanical point of view, is *hybridization of throws*, this phenomenon is connected with the energy saving for some specific throwing techniques, this means that several lever techniques can be changed in mist throws *Lever plus Couple* with very little amendments or little movements variation applying couple system to lever principle. Fig 16

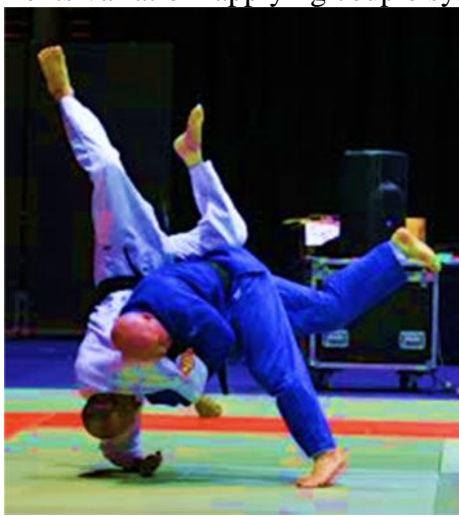

*Fig,15 Hybridization of Lever techniques Morote Seoi with Couple inserted*

*(Morote-Uchi Mata ?)*

Purist of Judo cry for involution in style, but I think that changing in fighting style is index of good health and life that flow some time in a better direction some time in a different one.

### 6) Male vs Female

It is well known that muscular strength between men and women is different; normally women have the 80%-85% of men's muscular strength in the legs and around the 60-70% in harms and grips. It is common knowledge that with strength training women's muscular force of harms could increase at near level of same weight men 85-90%. This information is not validated by scientific studies. The most advanced studies show us that the male –female difference is stable also after long hard training.

About grip strength and gender difference among men and untrained and well trained women, an interesting German study involving also well trained female Judo athletes it is performed with outstanding results.

Tis study performed by D.Leyk and coworkers in 2006 give us interesting results about real grip strength in women. [ 27]

In the following Diagrams there are three results from this interesting and not well known study.

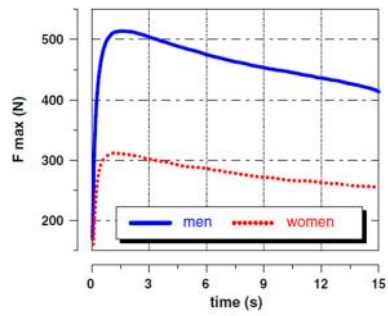

*Diag.1 Evolution in time of grip force in man and women*

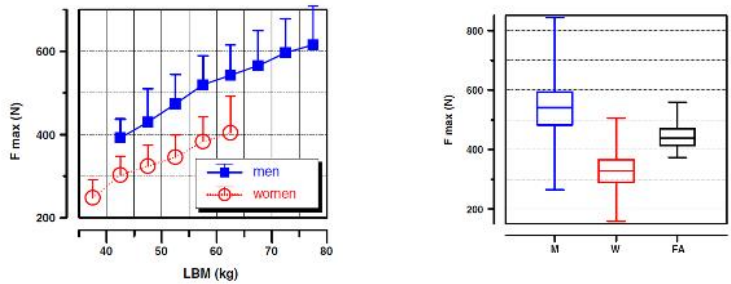

*Diag.2,3 Maximum hand grip Force of different Lean Body mass and Maximum hand grip force among non-trained men, women and female elite Judo athletes*

Then female elite athletes, also after long strength training have a lower hand grip strength than male untrained group. This situation lead to a different approach to competition; these results are also confirmed by a work proposed by Monteiro and coworkers [ 28 ] the following Diag. shows the power of elite Judokas female and male.

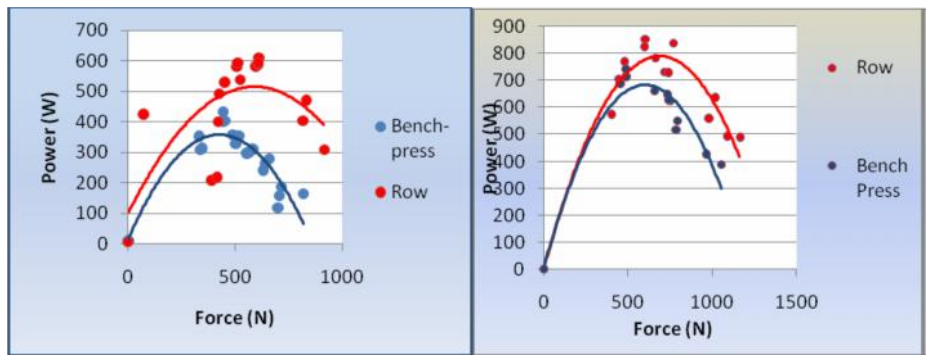

*Diag. 4,5  Female  and Male Power  Curve*

A thesis performed by Paulo Henrique Junqueira Hudson in 2007 on the world championship of the same year shows the following results:

| Throws with Ippon | Male    % | Female    % |
|---|---|---|
| *Te Waza* | 55 | 17 |
| *Ashi Waza* | 24 | 12 |
| *Sutemi Waza* | 14 | 12 |
| *Koshi Waza* | 3 | 0 |
| *Katame Waza* | 4 | 59 |

*Tab. 2  Example of difference in techniques application  between Male and Female (%) World Championship 2007*

These results are instantly connected to the previous ones' women apply the same percentage of (Ashi) legs techniques and (Sutemi) Sacrifices, but very few (Te) hand techniques and in contrary are more consistent and effective in (Ne Waza) groundwork.

Another big difference in women constitution is their greater joints' laxity and flexibility this is the reason that it is possible to see, high body's turn both during throwing actions and in Ne Waza, or long resistance with extreme angles in Kansetsu Waza application.

Also in attack rhythm there is difference Sterkowicz note in Olympic games of 1996 that " *Another characteristics feature of female athletes was the lower intensity of action during the attack and especially the frequency of penalties than in men who were better able to use the time of the fight.*"

Generally speaking women's grips application is more standard than men in percentage (Classic Ki Hon Kumi Kata left or right), but the increase of Russian gripping is also present form Nation to Nation.

More often application of throws is Innovative or Classic, very few Chaotic Forms are seen in women competition, but the percentage of Innovative variations is higher due to the body's flexibility of athletes. Connection Tachi Waza – Ne Waza, for Koshi Waza is very often linked to the application of Makikomi variation of throwing techniques as it shown in one of the following figures.

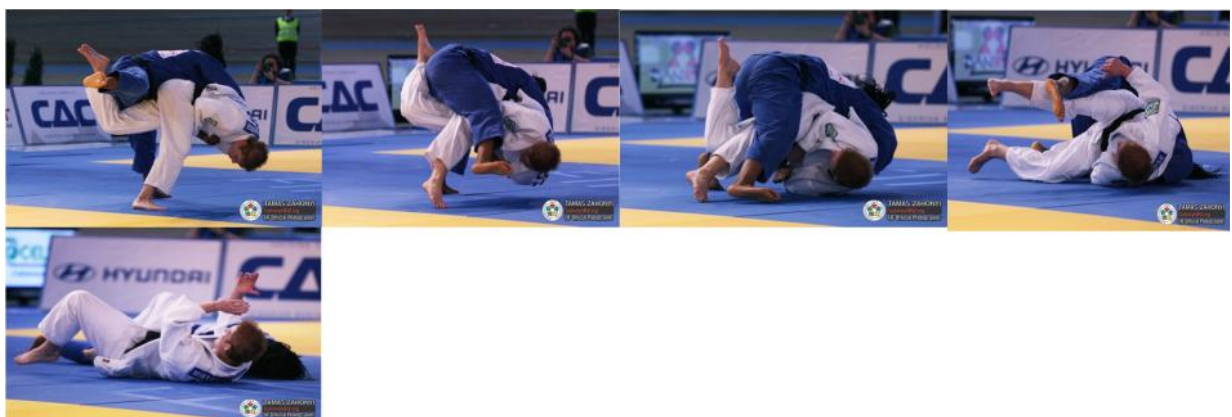

*Fig 16-20  High trunk flexibility in Innovative Uchi Mata*

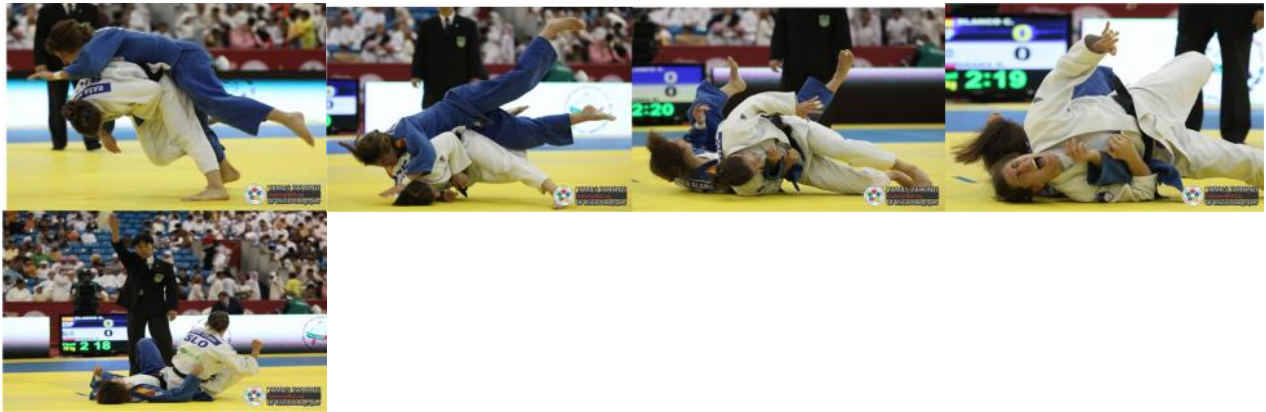

*Fig 21-25 Innovative Makikomi*

Normally, in women competitions, grip fight is less strength based, attack velocity is no as explosive as in men competition, and generally pace of contest is slower and it is possible to note that almost all female judo contests are smoothed confronted to male judo.
It is interesting to note that the poor presence of Chaotic Form of techniques in women games is directly connected to the natural and relative lack of strength both in hands and legs of female athlete's body structure.  Then women's judo, generally speaking,  remains more connected to Kodokan Judo as for grips preference, as for the form of throwing techniques applied (Classic or Innovative).
Now can be asked a common idea in West countries: Sport Judo is a male or female?
In the EU ranking (at 2007) based on gold medals the ranking is **1) RUS; 2) FRA; 3) NED; 4)GER.** [29]
However considering the grand total the result is   **1) RUS; 2) FRA; 3) GER; 4) UKR.**
But what is the importance of role of women judo in EJU and IJF?
If we are looking for the top four, highlighting the women role only, changes are introduced in EJU and bigger in the IJF previous rankings.

| Women         TOP   FOUR    EJU   NATIONS FOR  EJU  COMPETITIONS ||  |
|---|---|---|
| NATION | GOLD | GRAN TOTAL (all medals) |
| **FRA** | **61** | **199** |
| **RUS** | **45** | **193** |
| **NED** | **38** | **74** |
| **GER** | **32** | **85** |

*Tab.3 Top four EJU Nation women [29]*

| Women         TOP   FOUR    IJF  NATIONS FOR   ALL   COMPETITIONS ||  |
|---|---|---|
| NATION | GOLD | GRAN TOTAL (all medals) |
| **JAP** | **102** | **245** |
| **FRA** | **40** | **144** |
| **CHN** | **36** | **88** |
| **KOR** | **16** | **75** |

*Tab.4 Top four IJF Nation Women [29]*

If we see the male contribution to  Sport Judo, all previous ranking are speedy and totally changed, both in EJU and IJF, showing the big weight of women in the final ranking.As it is possible to see in the next two tables

| Men | TOP FOUR EJU NATIONS FOR EJU COMPETITIONS | |
|---|---|---|
| NATION | GOLD | GRAN TOTAL (all medals) |
| **RUS** | **92** | **261** |
| **GEO** | **50** | **189** |
| **FRA** | **30** | **128** |
| **AZE** | **24** | **85** |

*Tab.5 Top four EJU Nations Men [29]*

| Men | TOP FOUR IJF NATIONS FOR ALL COMPETITIONS | |
|---|---|---|
| NATION | GOLD | GRAN TOTAL (all medals) |
| **JAP** | **57** | **167** |
| **RUS** | **43** | **154** |
| **KOR** | **43** | **106** |
| **FRA** | **19** | **66** |

*Tab.6 Top Four IJF Nations Men [29]*

Then Sport Judo seems not a male prerogative, women play in it a very important role both in EJU and IJF.
The participation of women to Sport is a cultural problem in many countries, but we must also remember that, if the basic physical principles of throwing are the same for men and women athletes; Female Judo is totally different on the basis of physical and anthropometric differences, strength differences both in leg and more in arms and hands, or psychological differences and so on. All this means different way to think training, to approach technical preparation, or to choose the right class of applicable throwing techniques.
From the previous considerations, generally speaking about useful and effective throwing techniques, the question is obviously connected to athletes' body strength and morphology, then to their weight and preferred pace of competition, after to the physical skill and capability both of athletes and opponents.
But in general way women judo athletes prefer, Ashi Waza, Sutemi Waza and some Koshi waza in Makikomi evolution, all applied both in Classic and Innovative form.
Till now at EJU level three of the Four Top Nations give the just importance at Women Judo.
At world level only: Japan, France, China and Korea seem to have understood this lesson in right way.

### 7) *Fighting style evolution in time.*

In Biomechanical term, judo fighting style in high level competition, is a parameter changing in time.
That depends from a lot of external and internal variables like: Referee rules, changing in Competition Invariant, preferred pace of competition, changing in kumi kata, kind of throws utilized ( Classic, Innovative, Chaotic), type of Throwing Techniques ( Lever or Couple), training methodology, skill of athletes, high dynamical situations, and so on.
The fighting style is changed also because the influence of Japan culture decreased during time with the grows of foreign input into the world arena.
Till to the 70 before the entrance of soviet union in the Judo World, the Japanese style triumphed in the world arena, the most of fights were with classical Kumi Kata, right bodies position and high shifting Couple of Athletes velocity with high dynamic connected.
In the Table 7 [30] it is clearly visible the changing in fighting style from the Japan Idea: timing skill preference, fast Couple techniques that are useful with high shifting velocity, importance in Kaeshi and very low application of ne waza, with the more cautious fighting behavior after the 69 into the International Championships : Increase of Lever techniques importance, that means more quiet pace in competition, preference in coordinative skill, cautious grips position and very high ne waza contribution .

| Japanese competitions from 1929 to 1971 | | International Championships From 1969 to 1983 | |
|---|---|---|---|
| | % of Ippon | | % of Ippon |
| Uchi Mata | 16.9 | Seoi Nage | 9,6 |
| O Soto Gari | 12 | Uchi Mata | 9 |
| Harai Goshi | 8 | Harai Goshi | 6.8 |
| Seoi Nage | 7.5 | O Soto Gari | 6.6 |
| Tsuri Komi Goshi | 5.5 | Tai Otoshi | 4.6 |
| Tai Otoshi | 4.6 | Ne Waza | 38.1 |
| O Uchi Gari | 4.1 | | |
| Kaeshi Waza | 5.6 | | |
| Ne Waza | 6.7 | | |
| Tot. Ippon 1896 | 70.9 | Tot. Ippon 789 | 74.5 |

*Tab. 7 Difference in fighting style from Japan and rest of world [30]*

In the table 8 elaboration from di Hernandez & Torres. [31] It is shown the difference in technical application between male and female at 2007 world championship

| **Throws with Ippon** | **Male %** | **Female %** |
|---|---|---|
| *Te Waza* | 55 | 17 |
| *Ashi Waza* | 18 | 13 |
| *Sutemi Waza* | 14 | 12 |
| *Koshi Waza* | 3 | 0 |
| *Ne Waza* | 4 | 59 |

*Tab8 Example of difference in techniques application between Male and Female (%) World Championship 2007*

Also this evaluation shows the changing in fighting style , between male and female athletes due to the already similar training methods between the two genders.
In the next table there is shown a comparison between the percentage of male and Female Ippon during the most important International competitions.
In the Olympic year 2008 and the subsequent 2009- Male increases their effectiveness of more than 6%
Female shows a constant production

| *Confront between Ippon % during the 2008 and 2009* | | |
|---|---|---|
| Gender | Competition | % |
| Male | 2008 | 53,63 |
| | 2009 | 56,95 |
| Female | 2008 | 51,16 |
| | 2009 | 52,03 |

*Tab.9 Comparison between Male and Female amount of Ippon 2008-2009*

In Biomechanical term female athletes preferred specifically, during the decade 1990 - 2000 Couple Techniques most of trunk leg group, Physical Lever Techniques with minimum arm applied with Makikomi final closure, and Maximum Arm Techniques ( Sutemi) most applied with body weight and legs, few with a big contribution of arms. Few Chaotic Form, had seen because, normally, they need to be applied by use of strong arms contribution.
For their flexibility they utilized a lot Ne Waza with preference for Osae Waza.

But the ever and ever more similar training methods between male and female utilized during these years changed also the female style, from Japanese Classical to more strong planning attitude.
For example we can see in the next three table the preferred techniques utilized by female athletes during the world championships 2003 and 2005 divided by different weight classes

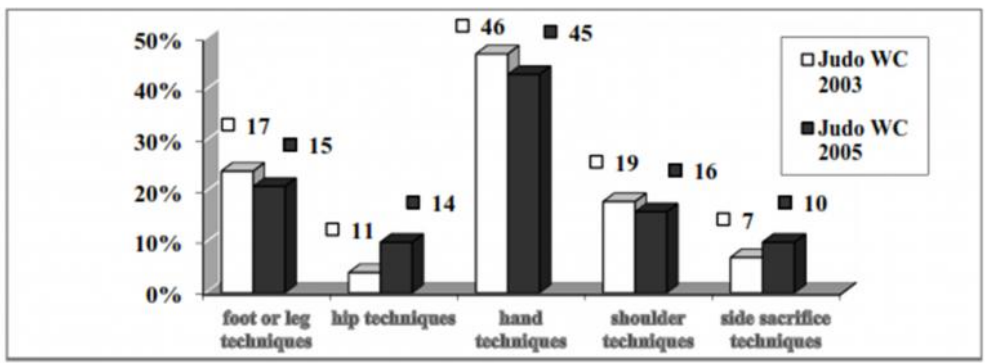

*Tab 10 light weight techniques utilization*

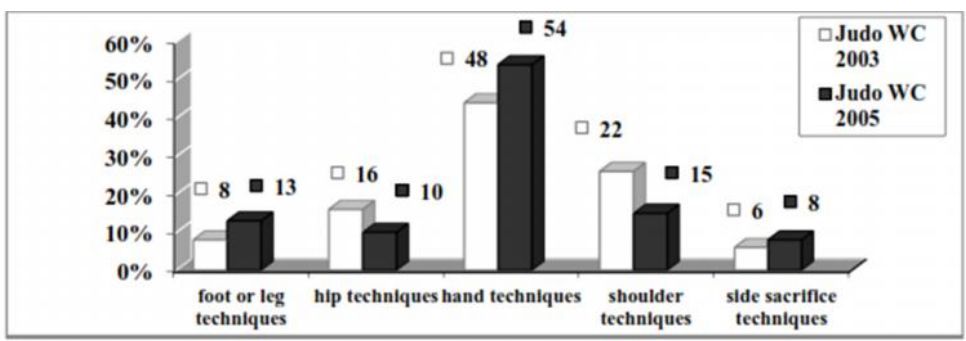

*Tab 11 middle weight techniques utilization*

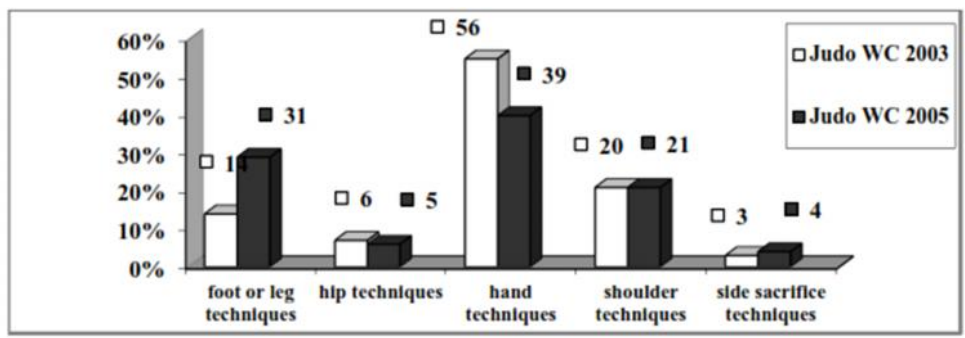

*Tab 12 heavy weight techniques utilization*

In this three tables we can see the fighting style changing for female high level athletes after 2000 year, divided for weight class during the world championship: very high use of hand techniques followed by shoulder techniques, very few sutemi and hip techniques [32]

Changing from timing skill to coordinative skill preferring Lever techniques than Couple techniques connected to timing skill.

In term of technical-tactical approach, analytical studies [ 32, 33, 34 ] show that the effectiveness of different types of attack is connected to determinate way to apply techniques.

Analysis produced on the 2010 world championship show ( see tabb. 13-14) the complex situations that are initiator of the Attack in high level judo competition and the percentage of utilization, from that it is easy to deduct that dynamics in high level competition is fundamental.

| Initiator of the attack | Situation immediataly before the attack | |
|---|---|---|
| Uke (attacks at the first) | Uke pushed and/or moved forward | ⬆ |
| | Uke pulled and/or moved backward | ⬇ |
| | Uke moved sideward or in acircle (pulled Tori on his way) | ↔ |
| | Uke moves after an attack the same way of Tsukuri back, or he remains in a fixed position or he stands up (in front of Tori) | O |
| Tori (attacks at the first) | Tori pulled and/or moved backward | ⬇ |
| | Tori pushed and/or moved forward | ⬆ |
| | Tori moved sideward or in acircle (pulled Tori on his way) | ↔ |
| Nobody | "Open" situation (the relation between push or pull is in ballance) | ✦ |
| Basic classes of Toris operation | Specific classes of Toris operation | |
| Tori is capitalizing a situation, who is initiated by Uke | Tori takes over the movement of Uke resp. whose push or pull action | ✚ |
| | Tori blocks or absorbs the movement of Uke and/or attacks against whose movement resp. push or pull attack | ✚ |
| Tori is capitalizing his own attack or an "open" situation | Tori self creates the situation and utilizes these directly | ➡ |
| | Tori self creates the situation and utilizes these indirectly | ✥ |
| | Tori takes over the "open" situation (at the first) | ◎ |

*Tab 13 WC 2010 Situation direct before the attack and classes of Tori's operation*

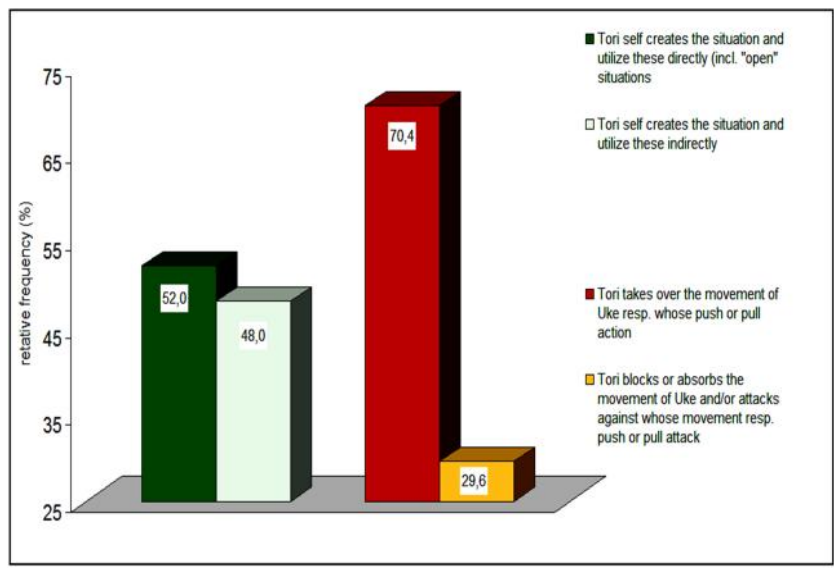

*Tab. 14 WC 2010 Relation ( in %) of generally and specifically classes of operation (Tori)*

Normally athletes attack in three main way **Direct attack, Action-Reaction strategy** and **Combination of Throws** this results show that the Action –reaction methods are the most useful, followed by Combination and Direct attack. In the next Diag. 5 the results are shown.

These three way to overcome the defensive capability of adversaries are the main practical way applied in real competition.
Also in the world Championship 2010 happen in Japan the following table shows the same use of the same different attack types.

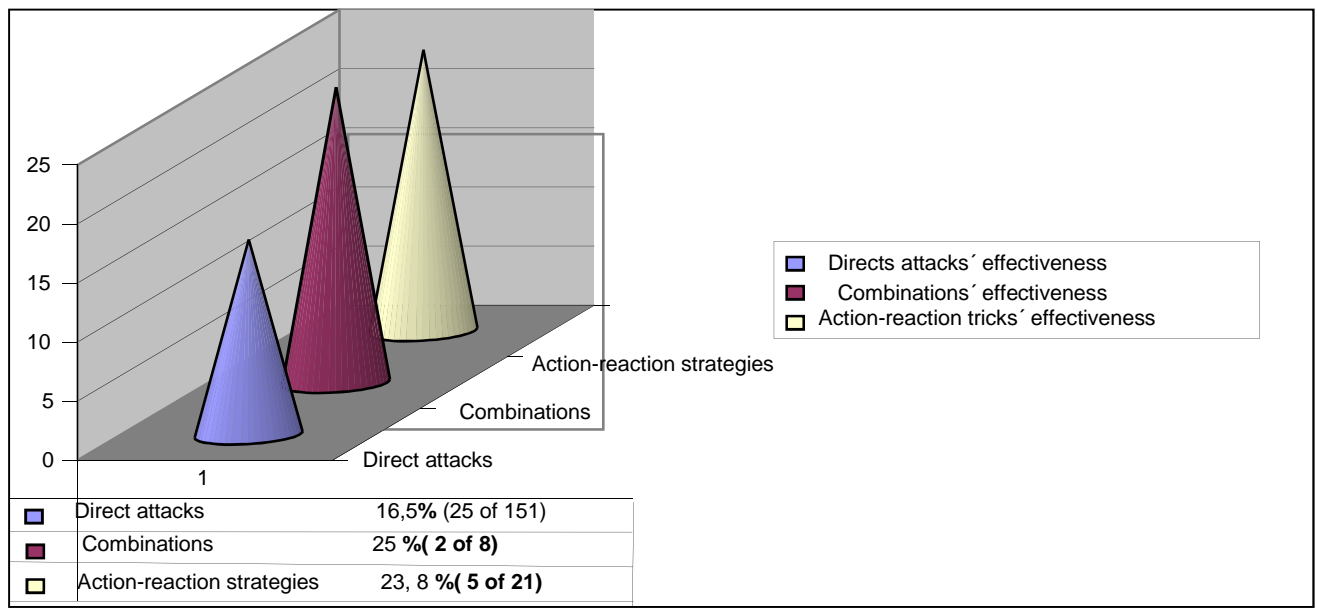

*Dig 6 Effectiveness of different types of attack*

| Judo Attack Type in World championship 2010 Japan | |
|---|---|
| Attack Tactics | % |
| Direct Attack | 42.2 |
| Action Reaction | 34.8 |
| Counter Attack | 16.3 |
| Combination | 8,1 |

*Tab.15 Effectiveness of different types of attack WC Japan 2010*

It is also interesting to see the Japanese fighting Style utilized during the same world championship in the following table clearly appears that Japanese athletes with little change prefer the dynamic style couple techniques more than lever, with an interesting introduction more sutemi to overcome the fighting style of foreign people

| Judo Skill of Japanese Male team WC 2010 Japan | |
|---|---|
| Techniques | % |
| Ashi Waza | 33.3 |
| Te Waza | 13.3 |
| Koshi Waza | 6.6 |
| Sutemi Waza | 33.3 |
| Ne Waza | 12.6 |

*Tab. 15 Judo Skill of Japanese Male team WC 2010 Japan*

With the slow motion of the fight can be understood the state of athlete's technical preparation and to take data for his technical improvement. This aspect of Judo Match Analysis is worldwide part of the today technical training and improvement.

Generally speaking the biomechanical problem of the technical improvement in Judo it is not only a problem of the athlete technical capability, but more often of the specific positional situation produced by the adversary in the couple system.

The effectiveness of attack (points) must be connected to the real application of techniques overcoming simultaneously the defensive system of the adversary.

The correct biomechanical approach is to consider the couple of athletes as a whole system in stable equilibrium into which the transition phase Kuzushi Tsukuri, depends strongly by the adversary position and action.

It is interesting to remember at this moment the main classes singled out in Japanese judo to define the right Tori action utilized to overcome the defensive standing of Uke, by Direct Attack:

1. *Tobi Komi* *(jumping in )*
2. *Mawarikomi (spinning in)*
3. *Hikidashi (pulling out)*
4. *Oikomi (dashing in)*
5. *Daki (to hug holding)*
6. *Debana (Thwarting the opponent)*
7. *Nidan Biki ( two stage pull)*
8. *Ashimoki (leg grab)*
9. *Sutemi ( body drop)*

However such interesting classification is practically useless for practical application, because in competition it is most useful to apply Action-Reaction tricks. Japanese term to call such strategies is ***Damashi Waza*** ( lit. feint techniques) or techniques applied after a feint ( body movements making-believe a technique ) more often a grip push-pull to produce a useful reaction of the adversary.

The third way to attack is combination : Japanese classification considers two kind of combination ***Renzoku*** and ***Renraku Waza***.

***Renzoku waza*** is a combination based on the attack continuous with the same type of throwing techniques, more often changing endlessly the attack angle.

***Renraku Waza*** is a combination based on the attack with different type of throwing techniques in different directions.

In judo the effectiveness of performance is grounded on synchronization, coordination and precision of movements and not on the force of movements, in high level competition, unbalance concept will change respect to classical view, it will be obtained with a more subtle concept changing the overall throwing aspect: the concept of ***"Breaking Symmetry".*** [35]

Then in high dynamic situations all judo throwing that are based on the two basic biomechanical tools, Couple and Lever will be applied by the following steps that reflect a continuous and fluent movement:

*Throws*
*Basic mechanic in high Level competition*

1. **First: breaking the adversary's Symmetry to slow down the opponent** (*i.e., starting the unbalancing action)*

2. **Second: timing,** *i.e.*, **applying the "***General Action Invariant***", with simultaneously overcoming the opponent's defensive grips resistance**

3. **Third: a sharp collision of bodies** (*i.e., the end of unbalance action)*

4. **Fourth:**

    A. **Application of "Couple of Forces" tool** (*i.e.*, **the type of throw), without any need of further unbalance action,**
        **Or**
    B. **Use of the appropriate "Specific Action Invariants", needing to increase unbalance action, stopping the adversary for a while, to apply the "Lever" tool** (*i.e.*, **the type of throw) in Classic or Innovative or New (Chaotic) way.**

The previous steps represent the simplest movements, which occur in a connected way and which are usually used to throw the opponent in high competitions. Very often though, far more complex situations can arise under real fighting conditions.

These complex situations which have evolved from the simple steps explained above depend on the combination of attack and defensive skills of both athletes.

However, the actual *Collision* step is very important for applying any real throwing technique.

This kind of collision is almost a rigid body collision [35].Nevertheless the use of rotational forces approach in *Advanced Kuzushi Concept* seems the most effective one's in term of score.

In the next figures it is shown the high level unbalance concept in dynamic competition.
 A ) application of a couple techniques without of further unbalance.
B) Action Reaction tricks, breaking symmetry stopping the adversary and collision with  Lever application.

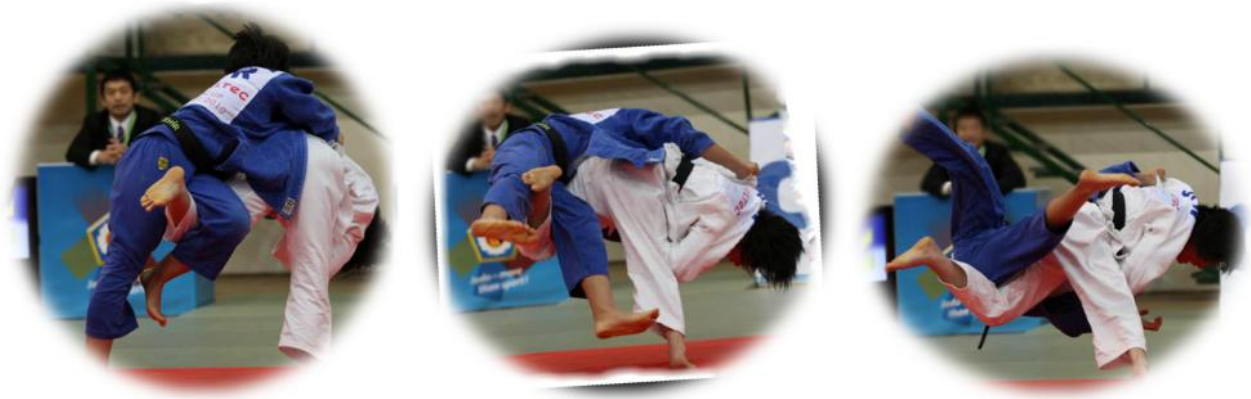

*Fig 26-28   A)   Diagonal application of  Trunk /Leg Couple without further unbalance*

*( Uke Rotates around his Center of Mass)*

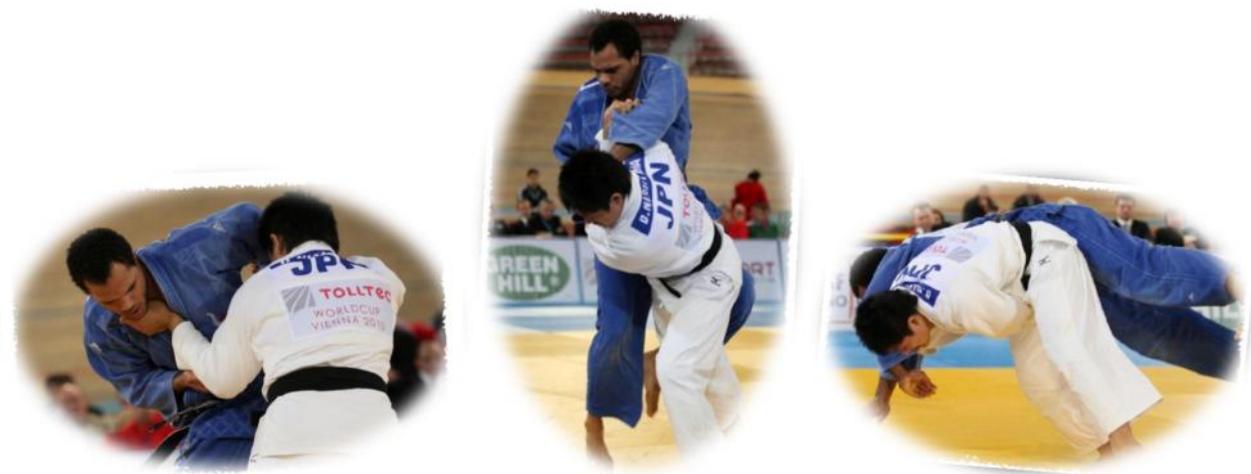

*Fig. 29-31  Breaking symmetry, Collision and Lever Application  with action reaction strategy*

*( Uke Center of Mass is translated in space)*

At the end it is possible to clarify the basic evolution of judo throws in time, with the changing attitude from old upright position and classic kumi kata till to the crouch body position and free kumi kata this more cautious attitude of Athletes Couple is connected to a slow shifting velocity and a quiet pace.

Attack velocity is today very high but the whole couple movement on the tatami is slow, then coordination, synchronization and precision based on a high fitness preparation substitute for timing and dynamical throws approach based on high shifting velocity.

In term of biomechanical classification of techniques this is translate by the augmented use of Lever Techniques in comparison to the Couple Techniques, with the counter ( Kaeshi Waza) reduction less than Uchi Mata Sukashi. From the variety application of techniques it is possible to see an increasing on different techniques used in direct attack and a decrease of different techniques dynamically connected in a combination, then in general term this means an increase of the total energy consumption in each contest.

The last Olympic show the same fighting style evolution as it is possible to see in the next two tables. [36]

*Table.16. Frequency of throw techniques used during Judo Olympic Tournaments (2012) by males and females from different weight categories*

| Technique of throws codes CODE | Total | Male | Female | Male groups/weight categories | | | | Female groups/weight categories | | | |
|---|---|---|---|---|---|---|---|---|---|---|---|
| | | | | *Group 1* | *Group 2* | *Group 3* | *Group 4* | *Group 1* | *Group 2* | *Group 3* | *Group 4* |
| LmaxA | 90 | 49 | 41 | 9 | 24 | 14 | 2[#] | 8 | 14 | 12 | 7 |
| LvA | 81 | 55 | 26* | 17 | 23** | 14** | 1** | 2 | 13 | 8 | 3 |
| CAL | 77 | 40 | 37 | 9 | 17 | 11 | 3 | 6 | 15 | 15[##] | 1[#] |
| CTL | 55 | 27 | 28 | 2 | 12 | 8 | 5* | 1 | 12 | 12 | 3 |
| LminA | 35 | 20 | 15 | 2 | 7 | 8 | 3 | 1 | 7 | 3 | 4 |
| LmidA | 11 | 3 | 8 | 0 | 1 | 1 | 1 | 0 | 4 | 2 | 2 |
| CA | 10 | 9 | 1 | 0 | 6 | 2 | 1 | 0 | 0 | 0 | 1 |
| Total | 359 | 203 | 156 | 39 | 90 | 58 | 16 | 18 | 65 | 52 | 21 |

*Legend for techniques code: Physical lever applied with max arm (LmaxA); Physical lever applied with variable arm (LvA); Couple of forces applied by arm or arms and leg (CAL); Couple of forces applied by trunk and legs (CTL); Physical lever applied with min arm (PLminA); Physical lever applied with middle arm (LmidA); Couple of forces applied by arms (CA). \*Significant difference between males and females, #Significant difference between group 1 and group 4, \*\* significant difference between group 1 and 2, 3, 4 groups, ## significant difference between groups 2nd and 3[rd]*

| *Throws Effectiveness In London Olympic 2012* | | |
|---|---|---|
| *Throws* | *Effectiveness Male %* | *Effectiveness Female %* |
| Seoi ( Ippon – Morote - Eri) | 14.8    (329) | 8.2    (222) |
| Uchi Mata | 9.2    (138) | 15    (143) |
| O Uchi Gari | 15    (53) | 24    (49) |
| Ko Uchi Gari | 12    (57) | 37    (35) |
| Tai Otoshi | 25    (36) | 23.8    (21) |
| Soto Makikomi | 10    (10) | 23.6    (17) |
| Tani Otoshi | 46    (13) | 50    (16) |
| Uchi Mata sukashi | 90    (10) | 100    (10) |
| *Couple* | *28.7* | *39* |
| *Lever* | *24* | *26.4* |

*Tab.17 Throws Effectiveness In London Olympic 2012*

The most important notations are the changing in fighting style that goes on in the same trend. This is confirmed by the growing use of Lever techniques , the slowing down of shifting velocity, the increase of quiet pace of contests that during the London Olympic flows into the growing of penalties for passivity.

Interesting are the results of the efficiency of throwing in contest, the data show that the effectiveness of throwing of females athletes is higher in Couple techniques and lower in lever techniques less than makikomi, all this depending from the body structure of female athletes, however the effectiveness mean of both throwing classes is already higher for female than for male athletes

### 8)  Advanced Mathematical tools for Match Analysis

In these last two paragraphs it is our goal to develop well throated pattern, utilizing mathematical tools to obtain useful information about high level Judo competitions. In the first part of this eighty paragraph we will face with the capability to obtain strategic information from the athletes pattern trajectories on the tatami during competition.

In the second one we deal with the problems : it is possible to evaluate the probability of a win in a contest or tournament?  And,  it is possible to forecast something of useful for coaching?

### 8.1) Mathematics for strategic evaluation

The motion patterns of the couple of athletes' system are a useful practical tool with hidden information inside.

The shifting patterns study could be source of very useful strategic data, but the price for extract the hidden information is a non-trivial mathematical analysis of these special time series.

In fact, for what concern the tracking trajectories one of the authors demonstrated 20 years ago, that the shifting paths of the couple of athletes COM projection must be considered belonging to the class of Brownian motions. In the fractional Brownian motion (fBm) approach, initially presented by Mandelbrot and van Ness in 1968, any time series can be considered a combination of deterministic and stochastic mechanisms.

The concept developed through fBm is, indeed, a generalization of the Einstein's work, which showed that a stochastic process is characterized by a linear relationship between mean square displacements  $<x^2>$ and increasing time intervals  $t$, in formula:

$$\langle x^2 \rangle = 2D\Delta t \qquad (1)$$

The general principle of the fBm framework is that the aspect of a trajectory, expressed as a function of time, may be calculated by a nonfinite integer or fractional space dimension, hence providing a quantitative measurement of evenness in the trajectory.
It is possible to write in mathematical form:

$$D_t^r [X(t)] - \frac{X(0)}{\Gamma(1-r)} t^{-r} = \xi(t) \tag{2}$$

The first term is a fractional derivative, the second is connected to the initial condition of the process, and the third is always the random force acting on the COM.
The fractional Brownian motion has the following covariance :

$$\langle x(t_1)x(t_2) \rangle = D_H \left[ t_1^{2H} + t_2^{2H} - |t_1 - t_2|^{2H} \right] = \Gamma(1-2H) \frac{\cos fH}{2fH} \left[ t_1^{2H} + t_2^{2H} - |t_1 - t_2|^{2H} \right] \tag{3}$$

In this case is important to know the mean square displacement of the point:

$$\langle [X(t) - X(0)]^2 \rangle = \frac{\langle \xi^2 \rangle}{(2r-1)\Gamma(r)^2} t^{2r-1} \propto t^{2H} \tag{4}$$

By this expression it is possible to understand that we are in presence of different diffusion processes, identified by the Hurst parameter.
In particular this parameter is time independent and it describes the fractional Brownian motion with anti-correlated samples for $0<H<1/2$ and with correlated samples for $½<H<1$.
If H is = to ½ we can speak of pure Brownian motion.
It is also very important that a fBm is connected to a Fractal based Poisson point processes, this special feature will be very useful and evilly utilized in the next paragraph, in order to find a right theoretical basis to evaluate victory probability and short term forecasting in a Judo match.
Davidsen & Schuster [37] drawn attention to a simple but plausible method for generating fractal-based point processes from ordinary Brownian motion.
Their construct resembles a conventional integrate-and- reset process but differs in that the threshold, rather than the integration rate, is taken to be a stochastic process.
This kind of behavior occurs in body's neurophysiology, for example, where ion-channel current fluctuations give rise to random threshold fluctuations. In the model considered by Davidsen & Schuster the rate remains fixed and the threshold process is taken to be ordinary Brownian motion. When the integrated state variable reaches the threshold, an output event is generated and the state variable is reset to a fixed value, as with a conventional integrated and reset process. It is also important to see the autocorrelation coefficient of fBm that, as well known, depends only by the time increment and not by the time function.
The autocorrelation coefficient for all sorts of fBm depends only from the ratio τ/t where τ=t'-t
.

$$\rho(\tau,t) = \frac{1}{2} \left( \left|\frac{t}{\tau}\right|^H + \left|\frac{\tau}{t}\right|^H - \left|\sqrt{\left|\frac{t}{\tau}\right|} - sng\left(\frac{\tau}{t}\right)\sqrt{\left|\frac{\tau}{t}\right|}\right|^{2H} \right) \tag{5}$$

For the special case $\tau=-t$ we have:

$$\rho(\tau,t) = \rho(-t,t) = 1 - 2^{2H-1} \tag{6}$$

we remember also that only for H=1/2 ( Regular Brownian Motion) autocorrelation coefficient for t and -t is independent, whereas fBm (t) and fBm(-t) are connected depending from the previous history. [38]

Athlete's Tracks (**Dromograms**) are the evolution in time of the couple of Athletes COM projection on the tratami area.

Normally in the old Match Analysis each technical action and throw was considered belonging to a class of Markovian System, this means that it depends by the previous instant only, without correlation with the past movements.

A more advanced mathematical approach let able to overcome this conceptual limitation and mathematical simplification.

As we have seen before, an important feature of fBm modeling, for each fighter, is the presence of long-term correlations between past and future increments.

This means that the system is not Markovian and then more similar to real situation.

It is interesting to note that the human paths produced by strategic thinking are very similar to track produced by inanimate elements.

This can be assessed by the scaling regimes.

***In this way a fighting path can show, if correctly analyzed, when the fighter have a specific fighting strategy or not (kind of random motion) during competition.***

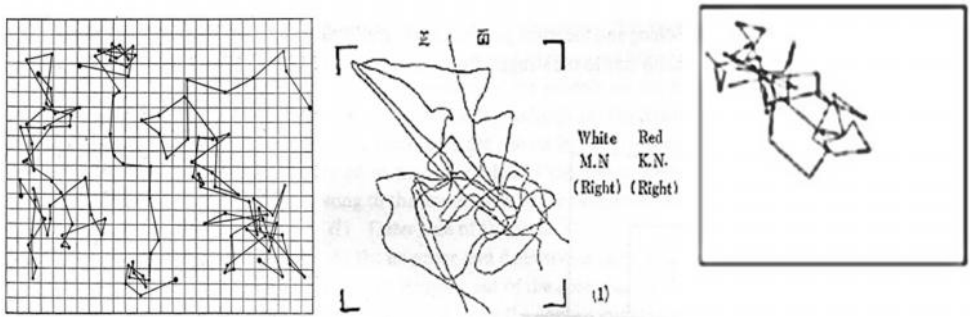

*Fig. 32. Three trajectories obtained by tracing a small grain of putty at intervals of 30 sec. Very similar to human shifting path CM projection on the tatami in competition. And Shifting path Computer simulation by fBm.*

For example a median value of 0.5 for *H* indicates that there is no correlation, suggesting that the trajectory displayed a random distribution (Brownian motion).

On the other hand, if *H* differs from 0.5, a positive (0.5 > *H*) or negative (H < 0.5) correlation with his fighting way can be inferred, indicating that a given part of initiative is under control.

Depending on how *H* is positioned, with respect to the median value 0.5, it can be inferred that the subject more or less controls the trajectory (and the fight evolution in time): the closer the regimes are to 0.5, the larger the contribution of stochastic processes (random attacks without strategy). In addition, depending on whether *H* is greater or less than the 0.5 thresholds, persistent (attacking) or antipersistent (defending) behavior can be revealed, respectively. In other words, if the CM projection at a certain time is displaced towards a given direction, the larger probability is that it drifts away in this direction (persistent attacking behavior) or in the contrary it retraces its steps in the opposite direction (antipersistent defensive behavior).

The trajectory obviously contain more information than the mean squared displacement. In particular one can measure the waiting time distribution from stalling events in the trajectories. For pronounced antipersistent processes immobilization events should be observed, i.e., for certain time spans neither coordinate should show significant variation (athlete stops the shifting action). [39] Due to the scale-free nature of fBm antipersistent these stops should span multiple time scales [40]. If such events occur they are indicative of the nature of the process. Absence of such features in shorter time series cannot necessarily rule out the fBm dynamics, in particular for H closer to one (ballistic motion) distinct stops are relatively rare events and possibly require very long time of fighting. Equality between these two probabilities ( H= ½) indicates that there is not presence of a defined strategy in fighting, like simple random motion or stochastic process.

This information obtained by a pure "mathematical lecture" of trajectories; can be enhanced adding to the previous advanced mathematical lecture other Biomechanical fighting information like Grips form, Competition Invariants, Action Invariants, Attack useful polygonal surface, Direction of displacement, Time and position of gripping action, Throws "loci", Length or Amount of displacement, Medium Speed, and Surface Area Utilization and so on.

It is possible, with this added information, to obtain a lot of useful strategic information structured as tree and to treat this tree of information with opportune Data Mining algorithms to obtain a categorization of potentially effective strategic connection among shifting trajectories and other Biomechanical fighting information.

Information, ordered by importance or effectiveness, is useful for coaching and athletes as well. This is one example of the more advanced information obtainable by this underestimated practical tool: Athletes' shifting patterns.

Our analysis started from the microscopic approach to COM motion in the space[41], that is fBm, from that derives the connection to the fBm described by the perpendicular of Athlete COM on the mat , it easy to induce that the motion of the Couple of Athletes COM perpendicular is again a fBm at microscopic level of fluctuation but was also proved that scaling at macroscopic level the motion is always Brownian.

In support of this connected reasoning a very interesting theoretical demonstration was presented independently into a new paper [42]. In it is presented a physical Langevin-based theory (Newton connected, remembering that it is a formal derivation of a Langevin equation from classical mechanics), explaining the emergence and the pervasiveness of the 'fractional motions' like : Brownian motion, Levy motion, fractional Brownian motion and fractional Levy motion.

In the article a general form of "micro-level" Langevin dynamics, with infinite degrees of freedom, is presented. Scaling from the micro-level to the macro-level the many degrees of freedom are summarized in only two characteristic exponents: the Noah and the Joseph exponent and the aforementioned fractional motions emerge universally. The previous two exponents categorize the fractional motions and determine their statistical and topological properties.

This useful theory establishes a unified 'Langevin bedrock' to fractional motions that as we know they are the basic description of Judo shifting paths.

### 8.2) *Mathematics for Probability and Short Term Forecasting*

From the theoretical point of view, now born the question it is possible in practical principle to have the winning probability for an athlete or forecast the result of his competitions?

It is very hard to answer such kind of questions, however from the mathematical point of view the answer is yes we have specific mathematical tools to do it !

In effect the solution of the answer is easy if we hypothesize that the attack pattern or occurrence follows the Poisson distribution. [43]

As already demonstrate, athletes' couple motion is described by a fBm of COM's projection, it is also well known that the fBm is connected to well specific Fractal based Poisson point processes. When a phenomena occur at discrete times (or places) with the individual events identical.

A point process is a mathematical construction which represents these events as points in a space, and is stochastic, when associated with random phenomena like Judo throwing attack or judo gait steps, etc.

For a stochastic point process, the statistics of this set of points provide information about the underlying structure of the phenomenon under study.

A fractal stochastic point process results when these statistics exhibit power-law scaling, indicating that the represented phenomenon contains clusters of points over all time or length scales [42]. When these discrete point processes follow the Poisson distribution we can speaks about Fractal based Poisson Point Processes

The Poisson process has been shown to be an acceptable representation of a number of physical phenomena in judo; e.g., the attack waves or pattern, the order of feet steps, the random success of the attack, and random failures of attack.

Simply put, the Poisson process could be a good mathematical descriptive model of **completely random** attack patterns. [44]

Obviously the problem is focalized on the effectiveness of the mathematical model proposed, able to utilize the Match Analysis data collected. This model could be improved modulating the parameters with opportune weights fitting the Match Analysis experimental data for each athlete. In effect the model must fit the technical capabilities of athletes to shows the victory probabilities. In the following will be show the probability evaluation for attack success of an hypothetical athlete.

*FOR EXAMPLE:*

*What is the probability of success ( whatever positive results ) at third minutes in competition, of one hypothetical athlete with four tokui waza 2 right and 2 left and three technical weakness 1 right and 2 left and fitness 5/8, after eight real attacks ?*

The Poisson attacks' counting process N(t) is Poisson distributed with attack mean λ at time t and the probability of attack success satisfies the following equation :

$$P[N(t)=n] = \frac{(\lambda t)^n}{n!} e^{-\lambda t} \qquad n=0,1,2,3.... \; number\,of\,attacks \qquad (7)$$

Where λt is the mean attack rate at time t that could be connected to the athletes' judo technical capability in contest time, in such way:

$$\lambda = (TA\text{-}TF/(R+L)+ Fitness\,) \left[1 - e^{\left(1-\frac{T}{t}\right)}\right] \qquad (8)$$

TA  = Technical attack ( number of Tokui Waza ) =  R+1,6L  
TF =  Technical failures ( number of weak spots) = 0.5R+0,8L  
R= Right Tokui Waza number  
L= Left Tokui Waza number  
Fitness = number of success/number of previous contests  
t=    asked time of contest  

*Answer  P =  0.13 or 13%*

A more adaptive description must also paint the adversaries' scouting data with his technical abilities and failures. In the case of presence of adversary technical capability equation , it is possible to apply a bivariate Poisson Process ( between athletes A and B) which probability follows the formula:

$$P(A,B) = e^{-(\lambda_A + \lambda_B + \lambda_0)t} \frac{(\lambda_A t)^n}{n!} \frac{(\lambda_B t)^m}{m!} \sum_{i=1}^{\min A,B} \binom{n}{i}\binom{m}{i} i! \left(\frac{\lambda_0}{\lambda_A \lambda_B}\right)^i \qquad (9)$$

With parameters' easy meaning.

Different approach is connected to the short term forecasting of Judo action, in this case it is possible to discuss only in term of Ippon obtained.
The obvious reason is that there is not a so sophisticated and flexible mathematical tools that is able to differentiate with precision the forecasting among three kinds of results that are inter-connected like Judo refereeing evaluations : Yuko, Waza Ari and Ippon, for these reasons we consider only Ippon and Waza Ari results that are 90% of Ippon as the same results.
In the case of short term forecasting , with all the limits in this field , we must remember that we are in presence of Fractal based Poisson Point processes.
The best solution in our knowledge will be an ARMA model, ( Auto Regression and Moving Average ) taking the value in time of different Ippon or waza ari obtained by the athlete during a tournament ( the best will be with the same adversaries ). These data normally could be arranged as random non stationary time series, then their autocorrelation could be used to determine whether there is any pattern in it. [46]
Classically in a ARMA model the time series of attacks to be forecast is expressed as function of both values ; previous values of the series (autoregressive terms) and previous error values from forecasting ( the moving average terms) .
The final concern will be to choose the correct ARMA dimension to obtain a short term forecast of ippon data.

### 9) Conclusion

With the modern evolution of computer and video analysis techniques born from the motion analysis, Match Analysis becomes ever and ever a most powerful Ju Do coaching tool.
The core of the systems are, generally speaking, a deeper analysis of the fight and the segmentation of the more interesting frames connected to a big data base.
A lot of information are achievable by the study of video capture, not only by qualitative sportive evaluations but also stunningly by means of advanced mathematical evaluation and reasoning.
Various kind of information physiological, biomechanical, technical and strategic, are available into the match analysis field.
Till today few improvement have been developed because this tool is undervalued by coaches and the only main focus is, more often, adversary scouting to obtain a win.
However the potentialities of the Match Analysis tool are more and more wide.
With a right training program and data it is possible to evaluate a minimum limit for the energy consumption, and the effectiveness of the changing of training program in time.
High information about technical failures and evolution, pros and contra of techniques, etc.
But stunningly important and useful information can be developed also by a deep mathematical analysis of specific results some normally found in ordinary Match Analysis , some other specifically developed.
 For example in the second field ( tool to develop for strategic goal) Athlete's Tracks (*Dromograms*) are the evolution in time of the couple of Athletes COM projection on the tratami area.
A modern mathematical approach let able to overcome the conceptual limitation of the old performed Markovian approach.
Very important feature of fBm modeling, for each fighter shifting path, is the presence of long-term correlations between past and future elements. This can be assessed by the effective presence of the scaling regimes. But the scaling regime in shifting path underlines another important connection the phenomenon ( gait steps) could be connected to a fractal based Point process.

***<u>Then a shifting path can show, if correctly analyzed, when the fighter have a specific fighting strategy or not (random motion) during competition</u>***.

For example a median value of 0.5 for *H* indicates that there is no correlation, suggesting that the trajectory displayed a random distribution (Brownian motion).
On the other hand, if *H* differs from 0.5, a positive (0.5 > *H*) o r negative (H < 0.5) correlation with his fighting way can be inferred, indicating that a given part of initiative is under control. Depending on how *H* is positioned, with respect to the median value 0.5, it can be inferred that the subject more or less controls the trajectory (and the fight evolution in time): the closer the regimes are to 0.5, the larger the contribution of stochastic processes (random attacks without strategy). In addition, depending on whether *H* is greater or less than the 0.5 thresholds, persistent (attacking) or antipersistent (defending) behaviors can be revealed, if H is greater than 1 for example the athletes' tendency to take some stops as already shown in a previous paper.
Obviously this information obtained by a pure "mathematical lecture" of trajectories; can be enhanced adding to the previous advanced mathematical lecture other Biomechanical fighting information like Grips form, Competition Invariants, Action Invariants, Attack useful polygonal surface, Direction of displacement, Time and position of gripping action, Throws "loci", Length or Amount of displacement, Medium Speed, and Surface Area Utilization and so on.
It is possible, with this added information, to obtain a lot of useful strategic information structured as tree and to treat this tree of information with opportune Data Mining algorithms to obtain a categorization of potentially effective strategic connection among shifting trajectories and other Biomechanical fighting information.
More deductions can be inferred, on the basis of a right mathematical model, by the Match Analysis data.
 Describing more judo aspect ( like generation of shifting path , gait steps or pace, attack  wave, etc.) by a Fractal based Poisson point process or simply by a Poisson Point process it is possible to obtain information about order of magnitude of the probabilistic evaluation of the attack success at a determinate time in competition, or  the tentative short term forecasting of a victory in competition like probability to take Ippon by means of ARMA methodologies.
This information, ordered by importance, effectiveness and obtainability, are useful for coaching and athletes as well to build a proper competition strategy. These are some examples of the more advanced information obtainable by right mathematical modeling and applications to this useful tool: Ju Do Match Analysis

42) **Sacripanti** Napoli Tokyo Napoli *Lectio Magistralis Univ. Di Napoli Federico II oct. 2010*
43) **Eliazar & Shlesinger** Langevin unification of fractional motions . *J Phys. A: Math. Theor.* (2012) 45
44) **Yahav & Shmueli** On generating multivariate Poisson data in management science applications *Applied stochastic model in business and industry 2011*
45) **Karlis & Ntzoufras** Analysis of sports data by using bivariate Poisson models *The statistician 2003 52*
46) **Bouzas, Aguilera & Valderrama** Forecasting a class of doubly stochastic Poisson processes *Statistical paper 2002 Spring & Verlag*38